# Competition between Alkalide Characteristics and Nonlinear Optical Properties in OLi$_3$−M−Li$_3$O (M = Li, Na, and K) Complexes

Ambrish Kumar Srivastava and Neeraj Misra[*]

Department of Physics, University of Lucknow, Lucknow - 226007, Uttar Pradesh, India

[*]Corresponding author, E-mail: neerajmisra11@gmail.com

**Abstract**

Alkalides possess enhanced nonlinear optical (NLO) responses due to localization of excess electrons on alkali metals. We have proposed a new class of alkalides by sandwiching alkali atoms (M) between two $Li_3O$ superalkali clusters at MP2/6-311++G(d,p) level. We notice a competition between alkalide characteristics and NLO properties in $OLi_3-M-Li_3O$ (M = Li, Na and K) isomers. For instance, the atomic charge on M ($q_M$) in $D_{2h}$ structure is -0.58$e$ for M = Li and its first static mean hyperpolarizablity ($\beta_o$) is 1 a.u., but in $C_{2v}$ structure, $q_M$ = -0.12$e$ and $\beta_o$ = 3.4×10$^3$ a.u. More interestingly, the $\beta_o$ value for M = K ($C_{2v}$) increases to 1.9×10$^4$ a.u. in which $q_M$ = 0.24$e$. These findings may provide new insights into the design of alkalides, an unusual class of salts and consequently, lead to further researches in this direction.

1. Introduction

Alkali metals generally possess +1 oxidation state and form ionic salts such as $Li^+F^-$, $Na^+Cl^-$ etc. with appropriate anions. During 1970s, Dye et al. [1] discovered an unusual salt [Na(2,2,2-crypt)]$^+$Na$^-$, which contains Na$^-$ isolated with Na$^+$ by cryptand ligand. This created the first example of an alkalide, the salts in which the anionic sites are occupied by alkali metals, i.e., Li$^-$, Na$^-$, K$^-$, Rb$^-$ etc. In subsequent studies, they reported a series of alkalides [2-5]. However, the high reactivity of alkalide anion results in the reductive cleavage of covalent bonds (such as C-O bond in a cryptand), destabilizing most of the alkalides even at the room temperature. Consequently, the room temperature stable alkalides have been designed using amines instead of ether linkages in cryptand such as K$^+$(aza222)M$^-$ (M = Na, K) [6]. This unusual -1 oxidation state of the alkali metals provides alkalides powerful reducing properties and unique electro-optical properties such as enhanced nonlinear optical (NLO) behaviour. Chen et al. [7] have designed a series of alkalides Li$^+$(calix[4]pyrrole)M$^-$ (M = Li, Na, K) and discussed their NLO properties. They proposed that the NLO behaviour of these alkalides increases with the increase in the atomic number of alkali anion (M$^-$). The enhanced NLO properties of AdzH$^+$Na$^-$ have been accredited to its small transition energy and high oscillator strength [8]. Some inorganic alkalide have also been reported such as "inverse sodium hydride" (H$^+$Na$^-$) [9] and Li$^+$(NH$_3$)$_4$M$^-$ (M = alkali atoms) [10].

The species with lower ionization potential than alkalis, so called as superalkalis [11] have been widely studied [12-16] due to their possible applications in the design of novel compounds [17-26] with unique and interesting properties. For instance, they can form unusual compounds with aromaticity [19, 20], enhanced electro-optical properties [21-23, 26] and increased gas phase basicity [25, 26]. Sun et al. [27] have employed superalkali clusters to design novel alkalides with larger NLO properties. They replaced Li$^+$ cations in Li$^+$(calix[4]pyrrole)M$^-$ alkalides by some superalkali cations such as Li$_3^+$, M$_3$O$^+$ (M = Li, Na, and K). Subsequently, they doped $Li_3$ superalkali in $NH_3$ [28] and noticed the alkalide characteristics of $Li_3(NH_3)_n Na$ ($n = 1-4$) with considerably large hyperpolarizablities. There are two important points noticeable in the design of these alkalides. First, they need a stabilizing ligand and second, their NLO properties are closely associated with the alkalide characteristics. Recently, we have suggested a simple design of alkalides by sandwiching alkali atoms between two $Li_2F$ superalkali clusters [29]. These alkalides are stabilized by charge transfer from $Li_2F$ superalkalis to alkali atoms, without use of any ligand. In this study, we propose a new series of alkalides using two $Li_3O$ superalkalis. $Li_3O$ superalkali has been observed experimentally by Jensen [30, 31] and already been used to design a number of compounds [21, 25, 27]. The novel alkalides, $OLi_3-M-Li_3O$ (M = Li, Na, and K) reported here, do not require any ligand. Furthermore, we will show that there is an interesting contest between alkalide and NLO characteristics such that the former becomes significant by the suppression of the later and vice versa.

## 2. Computational details

All calculations on $OLi_3-M-Li_3O$ (M = Li, Na, and K) complexes were performed using *ab initio* second order Møller-Plesset (MP2) perturbation theory [32] with 6-311++G(d,p) basis set as implemented in GAUSSIAN 09 program [33] with the aid of GaussView 5.0 software [34]. This scheme has recently been used to study a variety of superalkali compounds [25, 26]. The calculated vibrational harmonic frequencies were found to be positive which ensure that the optimized structures correspond to true minima. The natural bonding orbital (NBO) scheme [35] has been employed to calculated atomic charges on M atoms. The NLO properties were studied by calculating the mean polarizability ($\alpha_o$) and first order static mean hyperpolarizability ($\beta_o$) using the finite-field approach [36] by numerical differentiation with an electric field magnitude of 0.001 a.u.

$$\alpha_o = \frac{1}{3}(\alpha_{xx} + \alpha_{yy} + \alpha_{zz})$$

$$\beta_o = \left(\beta_x^2 + \beta_y^2 + \beta_z^2\right)^{1/2}; \beta_i = \frac{3}{5}\left(\beta_{iii} + \beta_{ijj} + \beta_{ikk}\right); (i, j, k = x, y, z)$$

### 3. Results and discussion

#### 3.1. Structures and bonding

In order to find the equilibrium structures, we have placed M atoms between two $Li_3O$ superalkali clusters along the axis passing through centre of $Li_3O$ as well as perpendicular to this axis. The geometry optimization leads to two planar structures of $OLi_3-M-Li_3O$ complexes, $D_{2h}$ and $C_{2v}$ for M = Li and Na as shown in Fig. 1. In $D_{2h}$ isomer, M atom is collinear with central O atoms of $Li_3O$ but inclined at about 45° in $C_{2v}$ structure. Table 1 lists various parameters associated with these structures. The $D_{2h}$ structure is 0.34 eV and 0.32 eV higher in energy than $C_{2v}$ for M = Li and Na, respectively. As mentioned earlier, these structures are planar in which both $Li_3O$ superalkalis and M atoms lie in the same plane. This is in contrast to $FLi_2-M-Li_2F$ in which $Li_2F$ superalkalis are perpendicular to each other [29]. For M = K, we do not find $D_{2h}$ structure of $OLi_3-M-Li_3O$ complex but only $C_{2v}$ structure. The bond-lengths of $OLi_3-M-Li_3O$ complexes are also listed in Table 1. One can see that the distance between Li and M atoms increases with the increase in the atomic number of M. For instance, Li−M bond-lengths are 3.23 and 3.37 Å (M = Li) which increase to 3.83 and 4.15 Å (M = K), respectively for $C_{2v}$ structure of $OLi_3-M-Li_3O$. On the contrary, Li−O bond lengths of $Li_3O$ in $OLi_3-M-Li_3O$ complexes are approximately same irrespective of M.

In order to explore the chemical bonding in $OLi_3-M-Li_3O$ complexes, we have performed quantum theory of atoms in molecule (QTAIM) [37, 38] analyses. The molecular graphs of $OLi_3-M-Li_3O$ calculated by QTAIM are displayed in Fig. 2 for M = Na. One can note that M interacts to $Li_3O$ superalkalis *via* O atoms in $D_{2h}$ structure whereas *via* Li atoms in $C_{2v}$ structure. These interactions can be classified by some topological parameters at the bond critical point (BCP) such as electron density ($\rho$) and its Laplacian ($\nabla^2\rho$). According to QTAIM, $\nabla^2\rho < 0$ for covalent bonding and ionic bonding is characterized by $\nabla^2\rho > 0$. For instance, $\rho$ and $\nabla^2\rho$ values for Na−O bonds in OLi$_3$−Na−Li$_3$O ($D_{2h}$) are 0.05027 a.u. and 0.41912 a.u., respectively, which suggest that the interactions between Na atom and OLi$_3$ moieties are essentially ionic. This bond is, however, weaker than corresponding Li−O bonds in OLi$_3$−Li−Li$_3$O ($D_{2h}$) as the bond lengths of M−O (M = Na) 4.51 Å are larger than 4.34 Å (M = Li). In OLi$_3$−Na−Li$_3$O ($C_{2v}$), $\rho$ and $\nabla^2\rho$ values are 0.00600 a.u. and -0.00002 a.u. for Na−Li, and 0.00771 a.u. and -0.00238 a.u. for Li−Li, respectively. The Li−Li bond length (3.34−3.35 Å) is not affected by the atomic number of M. Therefore, $C_{2v}$ structure is stabilized by not only the covalent interactions between M and Li$_3$O superalkalis but also between Li$_3$O superalkalis. This additional interaction provides significant stabilization in $C_{2v}$ structure of OLi$_3$−M−Li$_3$O complexes, which is evident from their relative energies with respect to $D_{2h}$ structure (see Table 1).

### 3.2. Alkalide characteristics

The NBO charges on M atoms, $q_M$ are listed in Table 2. One can see that all $q_M$ values are negative except for M = K. Therefore, OLi$_3$−M−Li$_3$O complexes are stabilized by the charge transfer from Li$_3$O superalkalis to M atoms. The charge transfer decreases with the increase in atomic number of M, i.e., IPs of M atoms. In particular, OLi$_3$−M−Li$_3$O ($D_{2h}$) structures possess $q_M$ values are -0.57$e$ for Li and -0.48$e$ for Na. These values are significantly large to establish the alkalide characteristics of OLi$_3$−M−Li$_3$O complexes. The alkalide behaviour of OLi$_3$−M−Li$_3$O is less pronounced than those of FLi$_2$−M−Li$_2$F complexes [29] in which NBO charge values on alkali anions are -0.69$e$ (M= Li) and -0.64$e$ (M = Na), but more effective than those of Li$^+$(calix[4]pyrrole)M$^-$ [7] possessing $q_M$ values of -0.33$e$ (M = Li) and -0.36$e$ (M= Na). In order to further explore the alkalide behavior of OLi$_3$−M−Li$_3$O complexes, we have plotted their molecular orbital surfaces in Fig. 3, including singly occupied molecular orbitals (SOMOs) and the lowest unoccupied molecular orbital (LUMO). One can see that the $\beta$-SOMO of OLi$_3$−M−Li$_3$O ($D_{2h}$) is localized on M atom, i.e., the excess electron cloud is concentrated on M resembling its anionic feature. In order to further reveal the electronic localization in OLi$_3$−M−Li$_3$O, we have plotted electron localization function (ELF) in Fig. 4 for M = Na. The ELF of $D_{2h}$ structure also suggests that the excess electron in the OLi$_3$−Na−Li$_3$O is located on the Na metal in accordance with the SOMO surface (Fig. 3). Furthermore, the ELF values for various regions indicate the existence of ionic interaction between Na and OLi$_3$ in $D_{2h}$ but covalent interaction between OLi$_3$ superalkalis in $C_{2v}$ structure, in addition to the ionic interaction as mentioned earlier.

The excess electron on alkali metal in alkalides is loosely bound due to very small electron affinity of alkali metal atoms. Consequently, alkalides possess lower IPs, even smaller than those of alkali metals. In Table 2, we have listed the calculated IPs of OLi$_3$−M−Li$_3$O complexes using Koopmans' theorem, which approximates the IP of a system to the negative energy eigenvalue of its HOMO. The calculated IPs of OLi$_3$−M−Li$_3$O are smaller than those of Li (~5 eV) which increase successively with the atomic number of M. This can be expected due to decrease in their alkalide characteristics, i.e., decrease in $q_M$ values. Similarly, the IPs of $C_{2v}$ structures are larger than those of $D_{2h}$. Thus, the increase in the alkalide nature results in the decrease in their stability and *vice versa*. In order to further explore this fact, we have calculated the difference in the energies of the HOMO and the LUMO, i.e., HOMO-LUMO energy gap ($E_{gap}$). It has been established that the systems with higher $E_{gap}$ values are electronically more stable and vice versa. The $E_{gap}$ of OLi$_3$−M−Li$_3$O complexes is also listed in Table 2. The $E_{gap}$ value increases with the increase in atomic number of M suggesting that the stability of $OLi_3-M-Li_3O$ is enhanced which is in accordance with the increase in their IP values and decrease in alkalide characteristics.

### 3.3. Nonlinear optical properties

The NLO properties of $OLi_3-M-Li_3O$ complexes have been explored by calculating mean polarizabilities ($\alpha_o$) and first order static hyperpolarizabilities ($\beta_o$) as listed in Table 3. One can note that the $\alpha_o$ of $OLi_3-M-Li_3O$ increases with the increase in the atomic number of M. Furthermore, the $\alpha_o$ of $D_{2h}$ structures is larger than their lower energy $C_{2v}$ structures, which reach to 1335 a.u. for $OLi_3-Na-Li_3O$. Note that $\alpha_o$ of $FLi_2-M-Li_2F$ is limited to 1151 a.u. for M = K [29] and those of $Li^+$(calix[4]pyrrole)$M^-$ and $Li^+(NH_3)_4Li^-$ have been reported to be 398-585 a.u. [7] and 498-879 a.u. [10], respectively. Therefore, mean polarizablity of these alkalides are smaller than those of $OLi_3-M-Li_3O$ complexes. The $\beta_o$ values of $OLi_3-M-Li_3O$ complexes show an interesting trend. It increases with the increase in the atomic number of M which has already been observed in case of $Li^+$(calix[4]pyrrole)$M^-$ alkalides [7]. More interestingly, $D_{2h}$ structures of $OLi_3-M-Li_3O$ possess vanishing $\beta_o$ values, although they possess significant alkalide characteristics for M = Li and Na. On the contrary, $C_{2v}$ structures possess significant $\beta_o$ values but not alkalide behaviour. For instance, $\beta_o$ value of $OLi_3-K-Li_3O$ is as high as $1.9\times10^4$ a.u., which is comparable to those of $Li^+$(calix[4]pyrrole)$K^-$[7] and those of other $OLi_3-M-Li_3O$ ($C_{2v}$) complexes are also in the same order ($\sim10^3$ a.u.) as of $Li^+$(calix[4]pyrrole)$M^-$[7]. Thus, there is a competition between alkalide characteristics and NLO properties of $OLi_3-M-Li_3O$ complexes. This feature is in contrast to many previous studies [7, 8], which demonstrated that large NLO responses results as a consequence of alkalide characteristics.

In order to explain this phenomenon, we have performed CIS/6-311++G(d, p) calculations on $OLi_3-M-Li_3O$ complexes. A qualitative estimation of $\beta_o$ may be obtained by following equation [39],

$$\beta_o \propto \frac{(\Delta\mu) f_o}{(\Delta E)^3}$$

where $\Delta\mu$ and $\Delta E$ are transition dipole moments and energies for crucial excited states, which is specified by the largest oscillator strengths ($f_o$). CIS calculated $\Delta E$, $f_o$ and $\Delta\mu$ values for crucial excited states are also listed in Table 3. The crucial excited states correspond to *β*-SOMO→LUMO+1 and *α*-SOMO→LUMO+1 transition for $D_{2h}$ and $C_{2v}$ structures, respectively. From Fig. 3, it is evident that the *β*-SOMO and *α*-SOMO of OLi$_3$−M−Li$_3$O is localized on M atom and between M and Li$_3$O moieties, respectively but LUMO+1 is delocalized over Li atoms of Li$_3$O moieties. Table 3 also lists the major contribution of electronic transition in crucial excited states. One can note that in $C_{2v}$ structures, $\Delta E$, $f_o$ and $\Delta\mu$ values increase along with the percentage contribution of the crucial transition, which results in the increase in $\beta_o$ for M = Li and Na. For M = K, both $f_o$ and $\Delta\mu$ further increase but $\Delta E$ decreases, which results in the enormous increase in the $\beta_o$ value. In $D_{2h}$ structures, although $f_o$ and $\Delta\mu$ values are larger than corresponding $C_{2v}$ structures and $\Delta E$ values are smaller, $\beta_o$ vanishes identically due to its symmetry about centre. This can be understood on the basis of the nature of electronic transition in crucial excited states. In $C_{2v}$ structures, *α*-SOMO→LUMO+1 transition corresponds to charge transfer from inner to outer Li atoms within Li$_3$O superalkali moieties, which results in larger $\beta_o$ value. On the contrary, *β*-SOMO→LUMO+1 transition in $D_{2h}$ structures leads to the delocalization of excess electron on M atom over outer Li atoms of Li$_3$O, which cause to vanish the $\beta_o$ due to symmetry.

4. **Final remarks and conclusions**

The structures of novel OLi$_3$−M−Li$_3$O complexes have been analyzed by placing alkali metal M (= Li, Na and K) between two Li$_3$O superalkalis. We have obtained two isomers, $D_{2h}$ and $C_{2v}$. The $D_{2h}$ isomer, which is higher in energy than $C_{2v}$, is stabilized by ionic interactions and $C_{2v}$ by covalent interactions. The NBO charges on M atoms in $D_{2h}$ isomer of OLi$_3$−M−Li$_3$O complexes are -0.58*e* and -0.47*e* for Li and Na, respectively, which suggest their alkalide characteristics. This is further supported by their lower ionization potentials and molecular orbital surface, showing the localization of excess electron on M atoms. However, their first static mean hyperpolarizability vanishes due to centrosymmetry. On the contrary, the $C_{2v}$ isomers have been shown to possess significant NLO properties, but not alkalide characteristics. Their first static mean hyperpolarizability is of order of 3.4−3.8×10$^3$ a.u. for M = Li, Na which increases abruptly to 1.9×10$^4$ a.u. CIS calculations have been performed to explain the NLO properties of OLi$_3$−M−Li$_3$O complexes. We have noticed that the isomers of OLi$_3$−M−Li$_3$O with appreciable alkalide nature fail to exhibit NLO responses and vice versa. These findings establish an interesting competition between alkalide behaviour and NLO properties.

**Acknowledgement**

A. K. Srivastava acknowledges Council of Scientific and Industrial Research (CSIR), New Delhi, India for a research fellowship [Grant No. 09/107(0359)/2012-EMR-I].

Table 1. Relative energy (Δ$E$) and bond-distances of OLi$_3$−M−Li$_3$O complexes at MP2/6-311++G(d,p) level.

| Complex | Sym | Δ$E$ (eV) | Li−M (Å) | O−M/Li−Li (Å) | Li−O (Å) |
|---|---|---|---|---|---|
| OLi$_3$−Li−Li$_3$O | $D_{2h}$ | 0.34 | 3.35 | 4.34 | 1.67, 1.75 |
|  | $C_{2v}$ | 0.00 | 3.23, 3.37 | 3.34 | 1.67, 1.75 |
| OLi$_3$−Na−Li$_3$O | $D_{2h}$ | 0.32 | 3.57 | 4.51 | 1.68, 1.74 |
|  | $C_{2v}$ | 0.00 | 3.49, 3.65 | 3.35 | 1.67, 1.75 |
| OLi$_3$−K−Li$_3$O | $C_{2v}$ |  | 3.83, 4.15 | 3.33 | 1.67, 1.74 |

Table 2. NBO charges ($q_M$), ionization potential (IP) and HOMO-LUMO gap ($E_{gap}$) of OLi$_3$−M−Li$_3$O complexes at MP2/6-311++G(d,p) level.

| Complex | Sym | $q_M$ (*e*) | IP (eV) | $E_{gap}$ (eV) |
|---|---|---|---|---|
| OLi$_3$−Li−Li$_3$O | $D_{2h}$ | -0.58 | 2.88 | 2.75 |
|  | $C_{2v}$ | -0.12 | 3.96 | 3.82 |
| OLi$_3$−Na−Li$_3$O | $D_{2h}$ | -0.47 | 3.07 | 2.92 |
|  | $C_{2v}$ | -0.01 | 3.65 | 3.49 |
| OLi$_3$−K−Li$_3$O | $C_{2v}$ | 0.24 | 3.63 | 3.47 |

Table 3. NLO parameters of OLi$_3$−M−Li$_3$O complexes obtained at MP2/6-311++G(d,p) level. CIS/6-311++G(d,p) results for crucial excited states are also listed.

| Complex | Sym | $\alpha_o$ | $\beta_o$ | $\Delta\mu$ | $f_o$ | $\Delta E$ | Crucial transition[*] |
|---|---|---|---|---|---|---|---|
| OLi$_3$−Li−Li$_3$O | $D_{2h}$ | 412.9 | 1 | 9.61 | 0.20 | 0.84 | $\beta$-S→L+1 (52%) |
| OLi$_3$−Li−Li$_3$O | $C_{2v}$ | 286.0 | 3436 | 5.34 | 0.22 | 1.65 | $\alpha$-S→L+1 (11%) |
| OLi$_3$−Na−Li$_3$O | $D_{2h}$ | 1335.5 | 0 | 14.76 | 0.30 | 0.82 | $\beta$-S →L+1 (55%) |
| OLi$_3$−Na−Li$_3$O | $C_{2v}$ | 523.0 | 3853 | 7.64 | 0.34 | 1.82 | $\alpha$-S →L+1 (31%) |
| OLi$_3$−K −Li$_3$O | $C_{2v}$ | 601.3 | 19406 | 9.29 | 0.41 | 1.80 | $\alpha$-S →L+1 (17%) |

[*]S = SOMO, L = LUMO

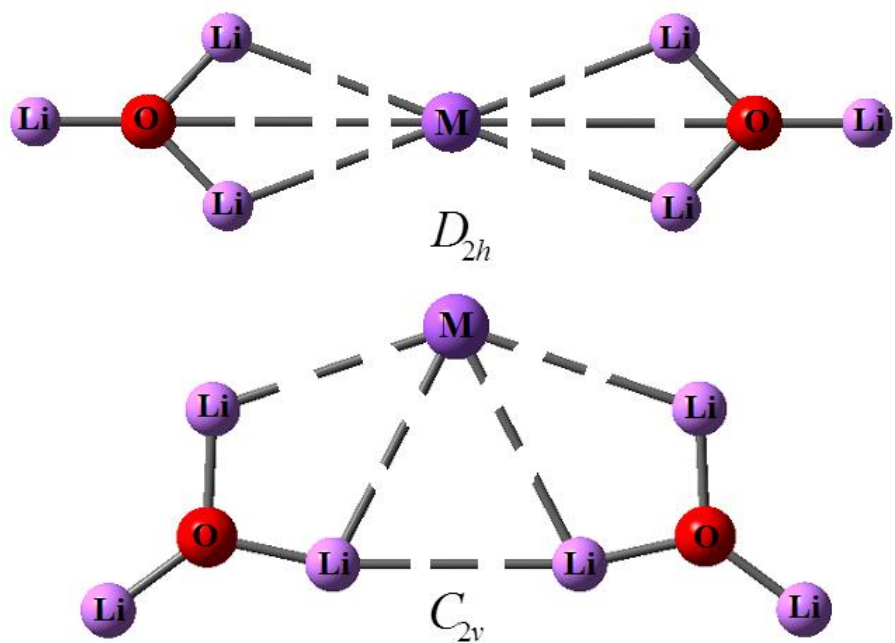

Fig. 1. Equilibrium structures of OLi$_3$−M−Li$_3$O (M = Li, Na and K) complexes at MP2/6-311++G(d,p) level.

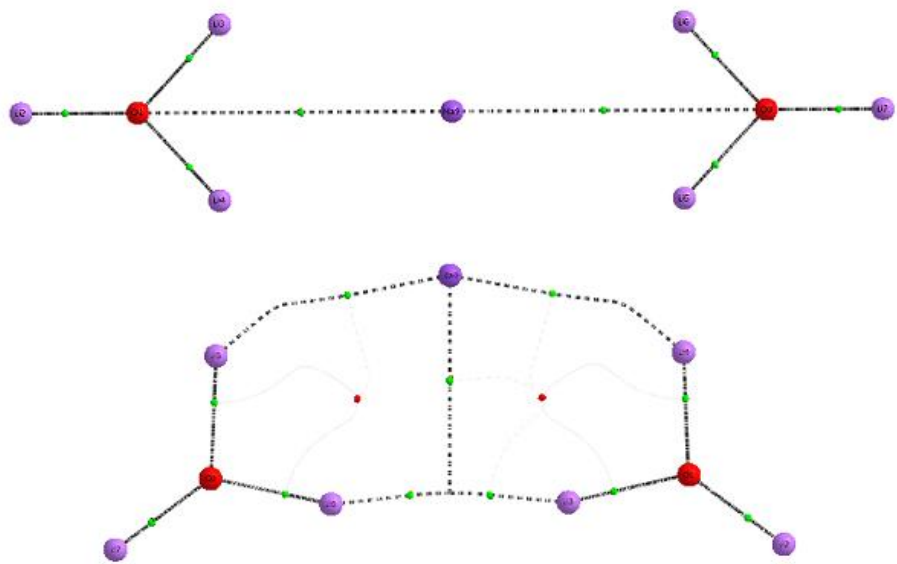

Fig. 2. Molecular graph of OLi$_3$−Na−Li$_3$O complexes calculated by QTAIM method.

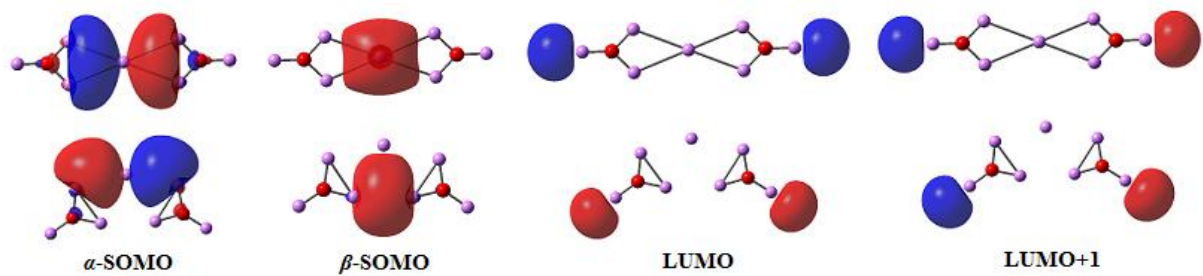

Fig. 3. Selected molecular orbital surfaces of OLi$_3$−M−Li$_3$O (M = Li, Na and K) complexes.

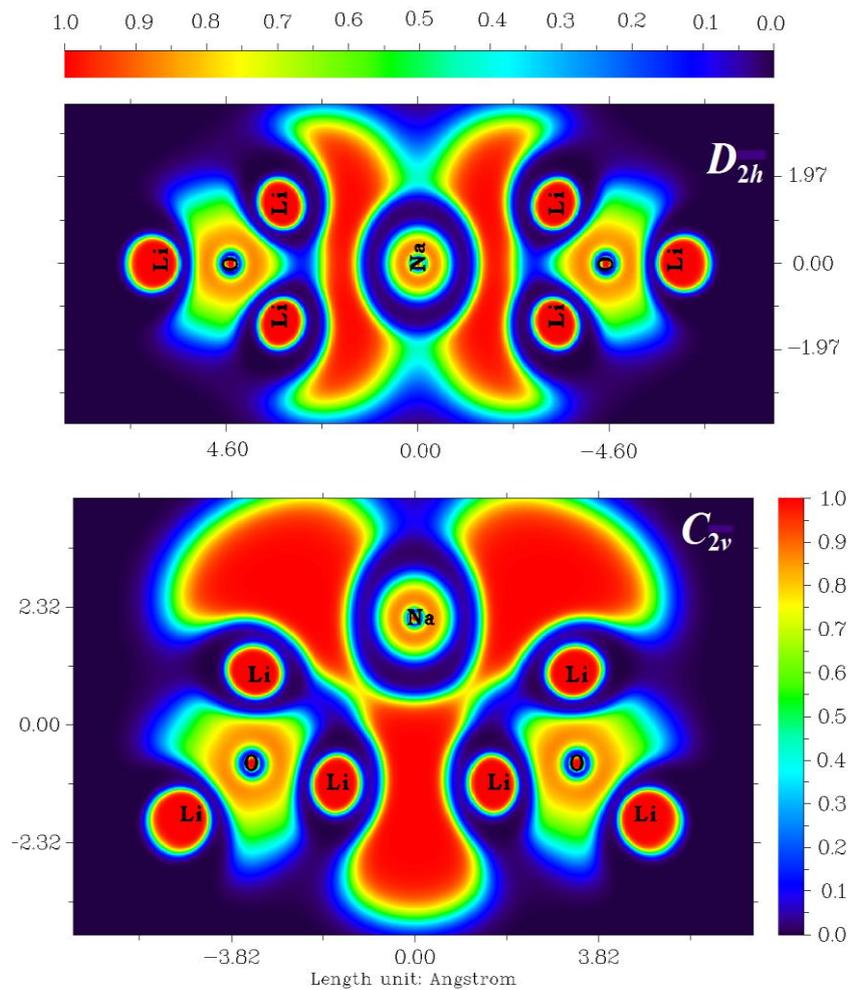

Fig. 4. The ELF map of OLi$_3$−Na−Li$_3$O complexes.